\documentclass[prb,superscriptaddress,twocolumn,floatfix]{revtex4}
\usepackage{graphicx}

\def\lcod{La$_2$CuO$_{4+\delta}$}

\def\lsco{La$_{2-x}$Sr$_x$CuO$_4$}

\def\ybco{YBa$_2$Cu$_3$O$_{6+x}$}
\def\lscoate{La$_{1.82}$Sr$_{0.18}$CuO$_4$}

\def\vqo{{\bf Q}_0}
\def\vqd{{\bf Q}_\delta}
\def\vqb{{\bf Q}_{\rm b}}

\begin{document}

\title{Evidence for an incommensurate magnetic resonance in
\lsco}
\author{J. M. Tranquada}
\affiliation{Physics Department,Brookhaven National Laboratory, Upton, NY 
11973-5000} 
\author{C. H. Lee}
\affiliation{National Institute of Advanced Industrial Science and
Technology, Tsukuba, Ibaraki 305-8568, Japan}
\author{K. Yamada}
\affiliation{Institute for Chemical Research, Kyoto University, Gokashou,
Uji, 611-0011 Kyoto, Japan}
\author{Y. S. Lee}
\affiliation{Department of Physics, Massachusetts Institute of
Technology, Cambride, MA}
\author{L. P. Regnault}
\author{H. M. R{\o}nnow}
\affiliation{CEA/Grenoble,D\'epartement de Recherche Fondamentale sur la
Mati\`ere Condens\'ee, 38054 Grenoble cedex 9, France}
\date{\today}
\begin{abstract} 
We study the effect of a magnetic field (applied along the $c$-axis) on
the low-energy, incommensurate magnetic fluctuations in superconducting
\lscoate.  The incommensurate peaks at 9~meV, which in zero-field were
previously shown to sharpen in {\bf q} on cooling below $T_c$ [T. E. Mason
{\it et al.}, Phys.\ Rev.\ Lett.\ {\bf 77}, 1604 (1996)], are found to
broaden in {\bf q} when a field of 10~T is applied.  The applied field
also causes scattered intensity to shift into the spin gap.  
We point out that the response at 9 meV, though occurring at
incommensurate wave vectors, is comparable to the
commensurate magnetic resonance observed at higher energies in other
cuprate superconductors.
\end{abstract}
\maketitle

\section{Introduction}

It has been observed in a variety of cuprate
superconductors\cite{regn95,dai01,fong99,meso00,he02} that the inelastic
magnetic scattering is enhanced below the superconducting transition
temperature, $T_c$, at a particular energy,
$E_r$, commonly referred to as the magnetic resonance energy.  The
``resonant'' magnetic scattering is found to be centered at the
antiferromagnetic wave vector and to have a rather narrow width in
energy.  The ratio $E_r/kT_c$ is observed to be in the range of 5 to 6.

One apparently anomalous system is \lsco.
To the best of our knowledge,\cite{kast98} no one has identified a
commensurate ``resonant'' response in this system by neutron scattering;
nevertheless, when certain theoretical interpretations of the optical
conductivity\cite{scha00} and angle-resolved photoemission\cite{esch00}
are applied to measurements on \lsco,\cite{sing01,zhou03} they seem to
imply a resonance at an energy of roughly 40 meV.  On the other hand,
Mason and coworkers\cite{maso96,lake99} found, for samples near optimum
doping, an enhancement of magnetic scattering below $T_c$ at {\it
incommensurate} wave vectors and occurring for energies centered at about
9 meV.  A concommitant narrowing in {\bf q} width was also observed.  It
seems possible that this effect corresponds to the commensurate resonance
seen in other cuprates.

To test the connection with the resonance phenomenon, it is desirable to
perform further characterizations.  One signature of the resonant magnetic
scattering in underdoped
\ybco\ is that the resonant scattering is reduced in amplitude by
application of a uniform magnetic field.\cite{dai00}  Here we study the
effect of a field on the incommensurate scattering in a slightly
overdoped crystal of La$_{1.82}$Sr$_{0.18}$CuO$_4$.  We find that, below
$T_c$, the applied field reduces the peak intensity of the incommensurate
scattering at 9 meV, thus providing support for associating the enhanced
incommensurate scattering with the commensurate resonance response found
in other cuprates.

There has also been considerable recent interest in the impact of an
applied field on magnetic scattering at lower energies.  In particular, an
applied field has been found to enhance elastic incommensurate
scattering in underdoped samples,\cite{kata00,lake02,khay02,khay03} and
to induce inelastic scattering within the spin gap of an optimally doped
sample.\cite{lake01}  For our slightly overdoped sample, it appears that
the field causes weight to shift into the gap from higher energy, causing
the frequency dependence to become more like that of the normal state
just above $T_c$.  These results are compared with a recent
study\cite{kimu03b} of Zn-doped La$_{1.85}$Sr$_{0.15}$CuO$_4$.

\section{Experimental Details}

The experiment was performed on
triple-axis spectrometer IN22 at the Institute Laue Langevin, which is
equipped with a vertically-focusing monochromator and a double-focusing
analyzer of pyrolytic graphite, using the (002) reflection.  No
collimators were used, but cadmium masks were placed as close as possible
to the sample (just outside of the magnet) to limit the beam size.  We
worked in fixed-$E_f$ mode, with $k_f=2.662$~\AA$^{-1}$ and a PG filter
after the sample.

The sample was an array of four crystals grown at Kyoto University,
and co-aligned in an aluminum holder.  The total crystal volume
was approximately 1.5 cm$^3$.  Magnetic susceptibility measurements
indicated that $T_c\approx 37$~K.  These crystals are similar to, but
distinct from, a sample of the same composition used in recent study of
the spin gap.\cite{lee03}  For the present sample, the
tetragonal-to-orthorhombic structural transition is at 118~K, whereas
the transition is at 111~K for the previous sample.  The higher
transition temperature corresponds to a slightly lower Sr content.

The crystals, oriented with the [001] direction
vertical, were mounted in a 12-T split-coil, vertical-field magnet. 
Thus, the applied field was along the $c$-axis, and we could study
scattering within the
$(hk0)$ zone.  (The [100] direction was aligned in the horizontal
scattering plane, but the [010] direction was tilted out of plane by
$\sim2^\circ$.)  We made use of an orthorhombic unit cell with
$a\approx b=5.316$~\AA.

For scans as a function of energy at fixed {\bf Q}, we should, in
principle, correct the intensities for energy-dependent counting-time
errors due to the presence of harmonics in the beam that reaches the
incident-beam monitor (see Chap.~4, Sec.~9, in
Ref.~\onlinecite{shir02}).  A correction factor is known for instruments
at the reactor face; however, IN22 is at the end of a thermal guide,
which should reduce the relative intensities of harmonics.  As we have
not measured the harmonic content of the incident beam, we are not able
to make the proper correction (which, at most, would involve a 20\%\
effect over the measured energy range).  This situation will have no
impact on the conclusions of our analysis, which focuses on the
variations of the inelastic signal with temperature and applied field;
however, this effect, together with the coarser resolution used here,
could be responsible for minor differences from the previous
study.\cite{lee03}

\section{Results}

\begin{figure}[t]
\centerline{\includegraphics[width=2.0in]{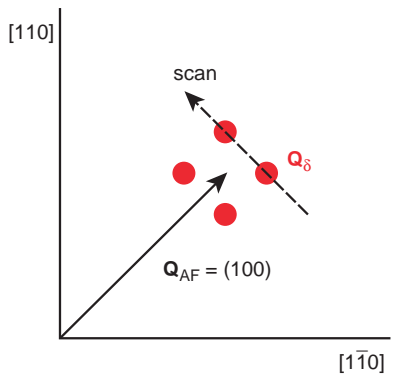}}
\medskip
\caption{(color online) Sketch of the $(h,k,0)$ zone of reciprocal space,
indicating the positions of the incommensurate magnetic wave vectors,
${\bf Q}_\delta$, which are split about the antiferromagnetic wave vector,
${\bf Q}_{\rm AF}$, denoted by the solid arrow.  The dashed
arrow indicates the path along which constant-energy scans were
performed, 
${\bf Q} = (1+\delta,k,0)$.}
\label{fg:recip} 
\end{figure}

The low-energy magnetic scattering in La$_{1.82}$Sr$_{0.18}$CuO$_4$ is
characterized by peaks at four incommensurate points about the
antiferromagnetic wave vector, ${\bf Q}_{\rm AF}$.  For a CuO$_2$ layer
with a square lattice, these peaks would be indexed as
$(\frac12\pm\delta,\frac12)$ and
$(\frac12,\frac12\pm\delta)$, with $\delta=0.13$.  In the orthorhombic
unit cell which we will use in this paper, the coordinates are rotated by
45$^\circ$, becoming $\vqd=(1+\delta,\pm\delta)$ and
$\vqd'=(1-\delta,\pm\delta)$, as shown in Fig.~\ref{fg:recip}. Because of
time constraints, most of the measurements involved measuring the
scattered intensity at the two peak positions $\vqd$ and at background
positions, $\vqb=(1+\delta,\pm0.4)$ and
$\vqo=(1+\delta,0)$, with a typical counting time of 15 min per point. 
(The actual measurements were done with $\delta=0.12$, rather than 0.13;
the difference is not significant for these measurements.) The background
measurements were found to be essentially independent of field, but
slightly temperature dependent (and, of course, energy dependent).  To
improve the statistics, the background measurements at each energy were
fit to a simple, monotonic function of temperature.  To obtain the net
intensity at $\vqd$, the fitted background was subtracted from the
average of the measurements at the two peak positions.

\begin{figure}[t]
\centerline{\includegraphics[width=2.8in]{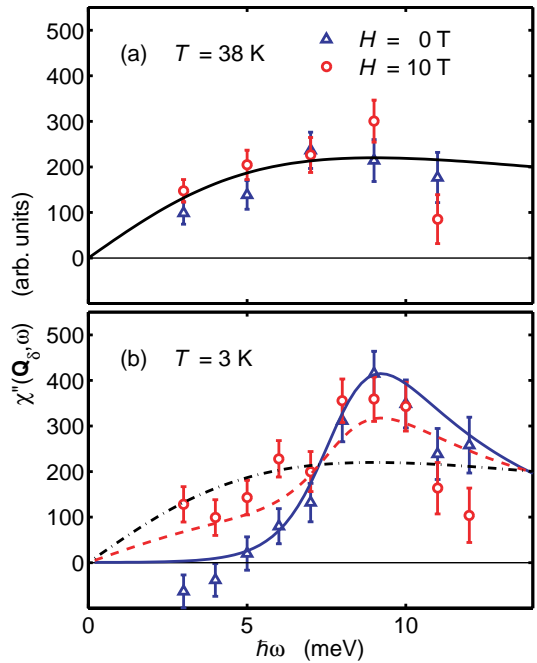}}
\medskip
\caption{(color online) Measurements of $\chi''(\vqd,\omega)$, in
arbitrary units, at (a) $T=38$~K, just above $T_c$, and
(b) $T=3$~K.  In both panels, the triangles (circles) denote measurements
at $H = 0$~T ($H=10$~T).  The lines through the data are explained in the
text.}
\label{fg:edep} 
\end{figure}

Figure~\ref{fg:edep} shows the energy dependence of the
imaginary part of the dynamic susceptibility, $\chi''$, at $\vqd$ measured
at temperatures of 3~K and 38~K for zero field and $H=10$~T. 
$\chi''$ was obtained by multiplying the net intensity by 
$1-\exp(-\hbar\omega/kT)$.  At $T\approx T_c$ [Fig.~\ref{fg:edep}(a)], the
differences in
$\chi''$ with and without a field are small, and probably due to
statistics.  The line through the data points corresponds to
\begin{equation}
  \chi_0'' = A_0{\hbar\omega\cdot \Gamma \over (\hbar\omega)^2+\Gamma^2},
\end{equation}
with $\Gamma = 9$~meV.  

\begin{figure}[t]
\centerline{\includegraphics[width=2.8in]{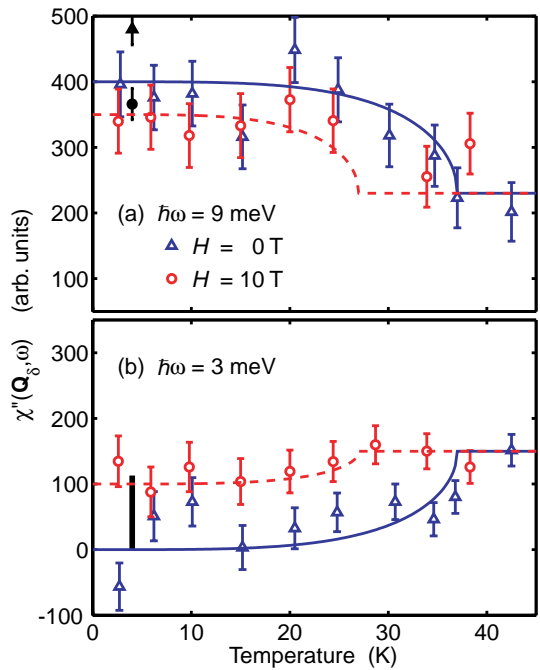}}
\medskip
\caption{(online color)  Temperature dependence of $\chi''(\vqd,\omega)$
measured at excitation energies of (a) 9 meV and (b) 3 meV.  In both
panels, the triangles (circles) denote measurements at $H = 0$~T
($H=10$~T).  The lines through the data are explained in the text.  The
two filled symbols in (a) correspond to the fits in Fig.~4(c),(d).  The
vertical bar in (b) corresponds to the fit of the intensity difference in
Fig.~4(b).}
\label{fg:tdep} 
\end{figure}

At $T\ll T_c$, Fig.~\ref{fg:edep}(b), we see a definite systematic
difference between zero field and 10-T measurements.  Applying the field
tends to introduce signal within the gap, and to decrease the
signal above the gap.  The solid curve through the zero-field data
corresponds to the phenomenological form
\begin{equation}
  \chi_{\rm sc}'' = A_1\chi_0'' [F_+(\omega)+F_-(\omega)]
  \left({\Delta_s\over\hbar\omega}\right)^2,
\end{equation}
where $\chi_0''$ (dot-dashed line) is from Eq.~(1) and
\begin{equation}
  F_\pm(\omega) = \tanh\left({\hbar\omega\pm\Delta_s\over\gamma}\right),
\end{equation}
with $\Delta_s=8$~meV, $\gamma=1.5$~meV, and $A_1=1.5$.  The dashed curve,
which roughly describes the in-field data, is given by
\begin{equation}
  \chi'' = 0.5\chi_{\rm sc}'' + 0.5\chi_0'',
\end{equation}
where $\chi_0''$ corresponds to the curve in Fig.~\ref{fg:edep}(a) at
38~K. The curves are intended to be suggestive guides to the eye, rather
than perfect fits to the data.

The temperature dependence of $\chi''$ at 3 meV and 9 meV is shown in
Fig.~\ref{fg:tdep}.  At 3 meV, the in-field data are
systematically finite and higher than the zero-field data for $T<T_c$.  At
9 meV, the in-field signal is reduced compared to zero-field.  The curves
are intended as suggestive guides to the eye, using a
BCS-like function, $\sqrt{1-(T/T_c)^4}$.  In zero field, the
measured $T_c$ is 37~K (solid lines), while for $H=10$~T, we estimate
$T_c=27$~K from the magnetization study of Li {\it et al}.\cite{li93}  

In their study of field effects on underdoped \ybco, Dai {\it et
al.}\cite{dai00} argued that the resonant response is a measure of
superconducting coherence.  The onset of coherent superconductivity is
reduced by the applied field, so that one would expect the onset of 9-meV
signal enhancement and 3-meV signal reduction to follow $T_c(H)$.  Our
measurements seem to be consistent with such a scenario; however, there
are insufficient data points at higher temperatures and the error bars
are too large to allow one to draw any firm conclusions regarding a
quantitative correlation with $T_c(H)$.

\begin{figure}[t]
\centerline{\includegraphics[width=3.2in]{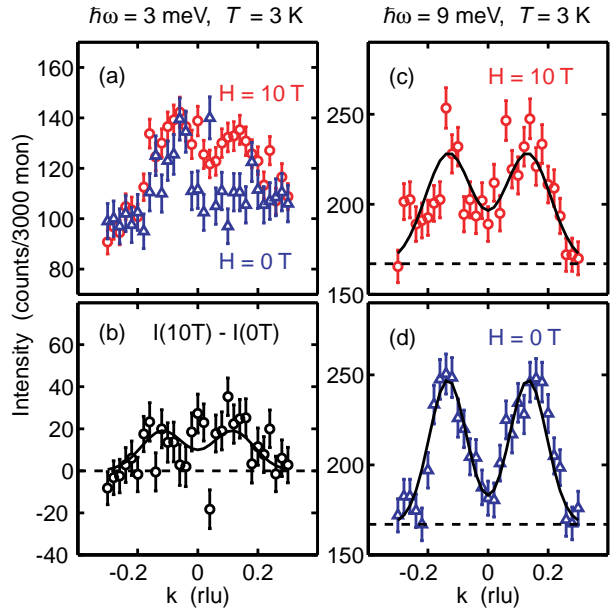}}
\medskip
\caption{(online color)  Constant-energy scans measured along the
direction indicated by the dashed arrow in Fig.~1.  All measurements are
at 3 K; scans in (a) and (b) are for $\hbar\omega=3$~meV, and (c) and (d)
are for $\hbar\omega=9$~meV.  In (a), (c), and (d), triangles (circles)
denote scans at $H=0$~T ($H=10$~T); (b) Shows difference between scans
from (a).  Lines are fits to symmetric Gaussian peaks, as discussed in the
text.}
\label{fg:scans} 
\end{figure}

Figure~\ref{fg:scans} shows constant-energy scans along 
${\bf Q} = (1+\delta,k)$ (see dashed line in Fig.~1) for
$\hbar\omega=3$~meV on the left and 9 meV on the right, all measured at
$T=3$~K.  The 3-meV scans have a strongly {\bf q}-dependent background
contribution that makes it difficult to analyze the raw data.  It is more
practical to look at the difference (high field $-$ zero field), shown in
(b).  The difference is consistent with a symmetric pair of broad peaks at
$k=\pm0.12(2)$.  The peak amplitude of 19(3)/3000 monitor counts is
consistent with the results in Fig.~\ref{fg:edep}(a) and Fig.~3(b) (see
the vertical bar in the latter), thus confirming the growth of low-energy
incommensurate scattering due to the presence of the field.

The 9-meV scans appear to have a more uniform background.  The curves
represent fits with symmetric Gaussian peaks.  In zero field, the peaks
are at $k=\pm0.134(3)$ with amplitude $= 80(4)$ and FWHM $=0.148(7)$; in
10~T the fit gives $k=\pm0.131(5)$, amplitude $=61(4)$, and FWHM
$=0.183(10)$.  Applying the field broadens the peaks and reduces the
amplitude; the amplitude change is consistent with
Figs.~\ref{fg:edep} and \ref{fg:tdep}(a) (see the filled symbols the
latter).  

In their study of La$_{1.86}$Sr$_{0.14}$CuO$_4$, Mason~{\it et
al.}\cite{maso96} observed at 9 meV an enhancement of intensity and a
narrowing in {\bf q} when cooling through $T_c$, which they discussed as
a coherence effect associated with superconductivity.  We find that
application of a 10-T field has the opposite effect: the magnetic
susceptibility is reduced, and the $q$-width is increased.  Again, this
seems to be consistent with a reduction in superconducting coherence due
to the field.

\section{Discussion}

\subsection{Resonance feature}

In our slightly overdoped sample, we find that application of a uniform
magnetic field parallel to the $c$ axis causes a reduction of $\chi''$ at
the energy of the peak ($\sim9$~meV).  The signal at this energy is
otherwise enhanced on cooling below $T_c$.  This behavior is reminiscent
of the field-induced decrease in the resonance peak observed\cite{dai00}
in underdoped \ybco, the main difference being that the response occurs
at an incommensurate, rather than commensurate, wave vector in \lsco.  
We note that in the original analysis of the zero-field enhancement of
the incommensurate signal, Mason {\it et al.}\cite{maso96}
suggested that the increase in signal below $T_c$ came from the
superposition of an extra contribution that is very narrow in $q$. 
Lacking a physical motivation for such a decomposition of the
excitations, we believe it is more reasonable to view the changes below
$T_c$ as a modification of the excitations that exist in the normal state.

In terms of the relative energy scale, the ratio $E_r/kT_c$ observed for
other cuprates\cite{regn95,dai01,fong99,meso00,he02} is found to lie in
the range of 5--6, as mentioned in the introduction.  If we identify
$E_r\approx9$~meV for our sample, then $E_r/kT_c\approx3$.   Relative to
$\Delta_0$, the maximum of the superconducting energy gap, $E_r$ is
observed to always be less than $2\Delta_0$, and generally not much
greater than $\sim\Delta_0$.  For \lsco\ with $x=0.18$,
$\Delta_0\approx10$~meV, based on
tunneling\cite{momo98} and Raman scattering studies,\cite{chen94} so
$E_r/\Delta_0$ is consistent with that for other systems.

Regarding energy scales, it is interesting to note that in a study of
\ybco\ with $x=0.51$ and $T_c=47$~K, Rossat-Mignod {\it et
al.}\cite{ross91} observed a spin gap of $\sim4$ meV in the
superconducting state together with an enhancement of $\chi''$ (with
respect to the normal state) peaked at $\sim7$ meV.  These energies are
comparable to those in our \lsco\ sample.  In more highly doped \ybco,
where the attention has tended to focus on the commensurate resonance
feature, we note that enhancements of $\chi''$ at incommensurate wave
vectors (for $E\ne E_r$) have also been
observed.\cite{bour00,dai98,rezn03}

There has been a variety of theoretical approaches to the magnetic
resonance and its energy and $q$ dependence.  From the perspective of
SO(5) theory, a model in which commensurate antiferromagnetism competes
with $d$-wave superconductivity, a magnetic resonance is predicted to
appear precisely at ${\bf Q}_{\rm AF}$.\cite{deml95,zhan97}  It
corresponds to a collective mode in the particle-particle channel, to
which neutrons cannot couple except in the superconducting state where
coupling is enabled by the coherent mixture of particles and holes in the
BCS condensate.  While the theory has been extended to include
(nontopological) stripes\cite{veil99} and dispersion of the
resonance,\cite{hu01} the commensurate resonance appears to remain a
central feature. 

One alternative is to attribute the resonance to an excitation of
antiferromagnetically-coupled Cu spins.\cite{morr98,aban99}  In the
normal state, interactions with the charge carriers cause the spin
fluctuations to be strongly damped, while fluctuations with energies
below $2\Delta_0$ become underdamped in the superconducting state.  Since
the $q$ dependence of the spin fluctuations is generally chosen to match
experiment in this approach, it can be either
commensurate\cite{morr98,aban99} or incommensurate.\cite{morr00}

The most common approach is to calculate the magnetic response of the
charge carriers themselves in the particle-hole channel, which is then
enhanced with the random phase 
approximation.\cite{lava94,mazi95,bulu96,dahm98,brin99,kao00,onuf02}
Whether the calculated fluctuations are commensurate or incommensurate
depends on the shape of the Fermi surface.\cite{dahm98,kao00,norm00} 
Using a model dispersion that gives a Fermi surface consistent with the
results of angle-resolved photoemission spectroscopy (ARPES) for \ybco\
and Bi$_2$Sr$_2$CaCu$_2$O$_{8+\delta}$ yields a commensurate resonance
peak.\cite{norm00}  

Calculations\cite{dahm98,kao00} for \lsco\ have
generally used parameters that give a Fermi surface that is closed about
${\bf k} = 0$, rather than about ${\bf k}={\bf Q}_{\rm AF}$ as in the
bilayer cuprates; however, it has been argued\cite{kao00} that the
differences in models are not essential for obtaining the normal-state
incommensurate structure in $\chi''$.  (We note that recent ARPES studies
indicate that the Fermi surface for optimally doped \lsco\ is actually
quite similar to that for the bilayer cuprates.\cite{dama03})  In any
case, a commensurate resonance feature is predicted\cite{dahm98,kao00} to
appear below $T_c$; in particular, Kao {\it et al.}\cite{kao00} predict
the resonance peak to occur at 15 meV.  While we must admit that we have
not pushed our measurements quite this high in energy, the maximum at 9
meV observed at an incommensurate wave vector does not appear to be
consistent with these calculations.

Some theorists have argued that there is a connection between the magnetic
resonance peak and certain anomalous features seen in ARPES measurements,
such as the ``peak-dip-hump'' structure\cite{aban99,esch00,bori03} and the
``kink'' in the quasiparticle dispersion.\cite{esch00,john01}  Eliashberg
theory has been used to make a connection between the resonance and
certain features in the optical conductivity.\cite{scha00}  (Theoretical
arguments against such connections have also been made.\cite{kee02}) 
Now, it happens that the same anomalous ``kink'' and optical conductivity
features identified for Bi$_2$Sr$_2$CaCu$_2$O$_{8+\delta}$ are also
observed for \lsco.\cite{zhou03,sing01}  To consistently interpret these
features in terms of the magnetic resonance, one would have to infer a
commensurate resonance at an energy of about 40 meV for \lsco.  Our
identification of the incommensurate 9-meV feature as the analog
of the resonant mode contradicts such an inference.

Finally, we note that the low-energy magnetic excitations in the normal
state of \lsco\ look very much like those observed\cite{tran99a,ito03} in
stripe-ordered La$_{1.48}$Nd$_{0.4}$Sr$_{0.12}$CuO$_4$.  In the latter
system, one interprets the incommensurate excitations as spin waves of the
magnetically ordered system.  The differences for \lsco\ can be understood
in terms of the fluctuations of a quantum-disordered system\cite{chub94}
with stripe correlations.\cite{kive03}  The magnetic excitations are
certainly sensitive to the charge fluctuations; after all, from the
stripe perspective, the incommensurability is the direct result of the
spatially inhomogeneous distribution of the doped
holes.\cite{zaan89,mach89,emer99}  The generation of a spin gap, together
with pairing of charge carriers, has been predicted based on a model that
assumes the existence of stripes.\cite{emer97}  One certainly expects
singlet-triplet excitations to appear above the spin gap.\cite{chub94}  A
model for the magnetic resonance based on incommensurate spin waves has
been proposed\cite{bati01}; however, a naive comparison with spin-wave
measurements in a stripe-ordered nickelate indicate that this model has
some shortcomings.\cite{bour03}

\subsection{Field-induced signal in the spin gap}

Neutron scattering experiments on underdoped \lsco\
(Refs.~\onlinecite{kata00,lake02}) and on \lcod\
(Refs.~\onlinecite{khay02,khay03}) have shown that application of a
magnetic field along the $c$ axis at temperatures less than $T_c$ can
induce or enhance spin-density-wave order.  While there have been a
number of proposals for the induced correlations in magnetic vortex cores,
\cite{arov97,fran02,chen02b,lee01b}  we believe that the most natural
explanation involves the pinning of charge and spin
stripes by
vortices.\cite{zhan02,polk02,kive02,zhu02,chen02c,taki03,ande03,kive03}
The observation that well developed charge and spin stripe order in
La$_{1.45}$Nd$_{0.4}$Sr$_{0.15}$CuO$_4$ is not affected by application
of a magnetic field is consistent with this picture.\cite{waki03}

In contrast to the underdoped regime, there is a spin-gap in the
superconducting state for optimally-doped
\lsco.\cite{yama95a,peti97,lake99}   The gap in the low-energy spin
fluctuations indicates that the spin stripes are further away from the
ordered state,\cite{polk02,kive02,kive03} so it is not surprising that an
applied magnetic field does not induce static correlations.  Instead,
Lake {\it et al.}\cite{lake01} showed, on a sample with $x=0.163$, that
applying a field induces a signal within the spin gap.  Our results are
generally consistent with theirs.  One difference is that they observed
an upturn in the low-energy (2.5 meV) in-field signal as the temperature
decreased below $\sim10$~K, whereas we did not see such an upturn 
in our slightly overdoped sample.

The application of the magnetic field in the superconducting state
introduces inhomogeneity associated with the vortices.  The
superconducting order parameter goes to zero at the center of each
vortex, and the area over which the order parameter is strongly depressed
is equal to $\pi\xi^2$, where $\xi$ is the superconducting coherence
length.  The areal fraction corresponding to the vortex cores is equal to
$H/H_{c2}$, where $H_{c2}$ is the field at which the sample becomes
completely filled by vortex cores.  The resistivity studies of Ando {\it
et al.}\cite{ando99b} indicate an $H_{c2}$ of approximately 55~T at 3~K
for \lsco\ with $x=0.17$, while the Nernst effect study of Wang {\it et
al.}\cite{wang02} suggests a low-temperature $H_{c2}$ of greater than
45~T for an $x=0.20$ sample.  Taking $H_{c2}\approx50$~T for our $x=0.18$
sample at 3~K, we find that, for our applied field of 10~T,
$H/H_{c2}\approx0.2$.  Thus, 20\%\ of the area is occupied by vortex
cores.

We expect that the magnetic scattering associated with the vortex cores
will be different from that due to the superconducting regions outside of
the cores.  We have seen that applying the magnetic field at 3~K causes
$\chi''$ to change so that it appears closer to the normal state.  At
10~T, the measurements can be roughly modeled as an average between
normal-state and zero-field-superconductor signals.  If the normal state
response came from just the vortex cores, then we would expect its weight
to be just 20\%\ instead of 50\%.  The larger normal-state response
indicates that it must come from regions about 2.5 times the area of the
vortex cores.  This result is consistent with an estimate\cite{esch01} for
the relative area in which the resonance is suppressed in
YBa$_2$Cu$_3$O$_{6.6}$.  The idea of a halo region extending beyond the
vortex core was suggested by the scanning tunneling microscopy study
of Bi$_2$Sr$_2$CaCu$_2$O$_{8+\delta}$ by Hoffman {\it et
al.}\cite{hoff02} and discussed by Zhang {\it et al.}\cite{zhan02}  A
much larger halo region is required to explain the neutron scattering
measurements\cite{kata00,lake02,khay02,khay03} of field-induced
spin-density-wave order in underdoped \lsco\ and \lcod.

We agree with Lake {\it et al.}\cite{lake01} that the magnetic field
induces a response that is closer to magnetic ordering; however, our
interpretation of that induced response differs somewhat from their's.  
They interpreted the induced response to be a mode within the spin gap,
with a peak energy much lower than the peak energy found in the normal
state above $T_c$.  Our results show that changes occur at higher
energies as well, so that the induced response is not restricted to the
spin-gap region. 

It is interesting to compare with a recent inelastic-neutron-scattering
study\cite{kimu03b} of Zn-doped \lsco.  In the muon-spin-rotation study
of Nachumi {\it et al.},\cite{nach96} it was deduced that each Zn dopant
reduces the superconducting carrier density by a fractional amount
corresponding to a relative area equal to that of a magnetic vortex
core.  One might then expect that the impact on spin excitations might be
similar to that from vortices.  Indeed, Kimura {\it et al.}\cite{kimu03b}
find that Zn-doping introduces a component of spin fluctuations that
extends into the spin gap of the undoped, $x=0.15$ parent material.  The
amount of signal within the spin gap grows with doping, and an elastic
component becomes detectable at a Zn concentration of 1.7\%.  At that
level of Zn, $T_c$ has been reduced from 37~K to 16~K.  That is a larger
change in $T_c$ than we are able to accomplish in our $x=0.18$ sample
with experimentally-achievable magnetic fields.  Of course, our sample is
on the metallic side of the insulator-to-metal crossover identified by
Boebinger {\it et al.}\cite{boeb96} using applied magnetic fields of
61~T, so that it seems unlikely that we would be able to induce static
spin stripe order in it simply by suppressing the superconductivity.

To avoid confusion, we should note that there are differences in the way
that we and Kimura {\it et al.}\cite{kimu03b} have presented the
inelastic results.  In presenting energy and temperature dependence, we
have shown $\chi''$ measured at a particular {\bf q} point, whereas
Kimura has plotted $\chi''$ integrated over {\bf q}.  Variations in
$q$-width of the inelastic peaks can cause the dependences of these
quantities on temperature, energy, etc.\ to be slightly different. 
Indeed, looking at the measurements at $\hbar\omega=9$~meV and $T=3$~K in
Fig.~4 (c,d), we see a drop in the peak intensity on applying the field;
however the peak area changes much less, since the width grows.

Vojta {\it et al.}\cite{vojt00} have shown that there is at least one
theoretical difference between the effects of a Zn dopant and a vortex:
substitution of a Zn atom for Cu effectively introduces a free spin. 
While these free spins can be detected by probes of the uniform spin
susceptibility,\cite{xiao90} it is not clear that they should play the
dominant role in the observed changes in inelastic scattering.  It seems
likely that the observed changes must come from a significant range about
each Zn, and that they involve a slowing of stripe fluctuations in the
vicinity of impurities, similar to the impact of vortices.

\section{Summary}

We have studied the effect of a magnetic field, applied parallel to the
$c$ axis, on the low-energy magnetic fluctuations in slightly-overdoped
La$_{1.82}$Sr$_{0.18}$CuO$_4$.  We observe that the enhancement of the
incommensurate intensity at 9 meV for $T<T_c$ is reduced when the field
is applied.  Based on this result, we identify the 9-meV peak as a
resonance feature in analogy with the commensurate resonance found in
other cuprates.  Field-induced signal is seen within the spin gap,
consistent with an earlier study, and indicating that the applied field,
which suppresses the superconductivity within vortex cores, also pushes
the magnetic correlations closer to a stripe-ordered state.  The
intensity of the in-gap signal indicates that it must come from a
region substantially larger than that of a vortex core.

\section*{Acknowledgements}

JMT is supported at Brookhaven by the U.S. Department of Energy's 
Office  of Science under Contract No.\ DE-AC02-98CH10886.  CHL's work at
AIST is supported by a Grant from the Ministry of Economy, Trade and
Industry of Japan.  KY receives support from the Japanese Ministry of
Education,  Culture, Sports and Science and Technology; Grant-in-Aid for
Scientific Research on Priority Areas, Scientific Research (B),
Encouragement of Young Scientists, and Creative Scientific Research. 
This study was supported in part by the U.S.-Japan Cooperative Research
Program on Neutron Scattering.


\begin{thebibliography}{79}
\expandafter\ifx\csname natexlab\endcsname\relax\def\natexlab#1{#1}\fi
\expandafter\ifx\csname bibnamefont\endcsname\relax
  \def\bibnamefont#1{#1}\fi
\expandafter\ifx\csname bibfnamefont\endcsname\relax
  \def\bibfnamefont#1{#1}\fi
\expandafter\ifx\csname citenamefont\endcsname\relax
  \def\citenamefont#1{#1}\fi
\expandafter\ifx\csname url\endcsname\relax
  \def\url#1{\texttt{#1}}\fi
\expandafter\ifx\csname urlprefix\endcsname\relax\def\urlprefix{URL }\fi
\providecommand{\bibinfo}[2]{#2}
\providecommand{\eprint}[2][]{\url{#2}}

\bibitem[{\citenamefont{Regnault et~al.}(1995)\citenamefont{Regnault, Bourges,
  Burlet, Henry, Rossat-Mignod, Sidis, and Vettier}}]{regn95}
\bibinfo{author}{\bibfnamefont{L.~P.} \bibnamefont{Regnault}},
  \bibinfo{author}{\bibfnamefont{P.}~\bibnamefont{Bourges}},
  \bibinfo{author}{\bibfnamefont{P.}~\bibnamefont{Burlet}},
  \bibinfo{author}{\bibfnamefont{J.~Y.} \bibnamefont{Henry}},
  \bibinfo{author}{\bibfnamefont{J.}~\bibnamefont{Rossat-Mignod}},
  \bibinfo{author}{\bibfnamefont{Y.}~\bibnamefont{Sidis}}, \bibnamefont{and}
  \bibinfo{author}{\bibfnamefont{C.}~\bibnamefont{Vettier}},
  \bibinfo{journal}{Physica B} \textbf{\bibinfo{volume}{213\&214}},
  \bibinfo{pages}{48} (\bibinfo{year}{1995}).

\bibitem[{\citenamefont{Dai et~al.}(2001)\citenamefont{Dai, Mook, Hunt, and
  Do\u{g}an}}]{dai01}
\bibinfo{author}{\bibfnamefont{P.}~\bibnamefont{Dai}},
  \bibinfo{author}{\bibfnamefont{H.~A.} \bibnamefont{Mook}},
  \bibinfo{author}{\bibfnamefont{R.~D.} \bibnamefont{Hunt}}, \bibnamefont{and}
  \bibinfo{author}{\bibfnamefont{F.}~\bibnamefont{Do\u{g}an}},
  \bibinfo{journal}{Phys. Rev. B} \textbf{\bibinfo{volume}{63}},
  \bibinfo{pages}{054525} (\bibinfo{year}{2001}).

\bibitem[{\citenamefont{Fong et~al.}(1999)\citenamefont{Fong, Bourges, Sidis,
  Regnault, Ivanov, Gu, Koshizuka, and Keimer}}]{fong99}
\bibinfo{author}{\bibfnamefont{H.~F.} \bibnamefont{Fong}},
  \bibinfo{author}{\bibfnamefont{P.}~\bibnamefont{Bourges}},
  \bibinfo{author}{\bibfnamefont{Y.}~\bibnamefont{Sidis}},
  \bibinfo{author}{\bibfnamefont{L.~P.} \bibnamefont{Regnault}},
  \bibinfo{author}{\bibfnamefont{A.}~\bibnamefont{Ivanov}},
  \bibinfo{author}{\bibfnamefont{G.~D.} \bibnamefont{Gu}},
  \bibinfo{author}{\bibfnamefont{N.}~\bibnamefont{Koshizuka}},
  \bibnamefont{and} \bibinfo{author}{\bibfnamefont{B.}~\bibnamefont{Keimer}},
  \bibinfo{journal}{Nature} \textbf{\bibinfo{volume}{398}},
  \bibinfo{pages}{588} (\bibinfo{year}{1999}).

\bibitem[{\citenamefont{Mesot et~al.}(2000)\citenamefont{Mesot, Metoki, Bohm,
  Hiess, and Kadowaki}}]{meso00}
\bibinfo{author}{\bibfnamefont{J.}~\bibnamefont{Mesot}},
  \bibinfo{author}{\bibfnamefont{N.}~\bibnamefont{Metoki}},
  \bibinfo{author}{\bibfnamefont{M.}~\bibnamefont{Bohm}},
  \bibinfo{author}{\bibfnamefont{A.}~\bibnamefont{Hiess}}, \bibnamefont{and}
  \bibinfo{author}{\bibfnamefont{K.}~\bibnamefont{Kadowaki}},
  \bibinfo{journal}{Physica C} \textbf{\bibinfo{volume}{341}},
  \bibinfo{pages}{2105} (\bibinfo{year}{2000}).

\bibitem[{\citenamefont{He et~al.}(2002)\citenamefont{He, Bourges, Sidis,
  Ulrich, Regnault, Pailh\`es, Berzigiarova, Kolesnikov, and Keimer}}]{he02}
\bibinfo{author}{\bibfnamefont{H.}~\bibnamefont{He}},
  \bibinfo{author}{\bibfnamefont{P.}~\bibnamefont{Bourges}},
  \bibinfo{author}{\bibfnamefont{Y.}~\bibnamefont{Sidis}},
  \bibinfo{author}{\bibfnamefont{C.}~\bibnamefont{Ulrich}},
  \bibinfo{author}{\bibfnamefont{L.~P.} \bibnamefont{Regnault}},
  \bibinfo{author}{\bibfnamefont{S.}~\bibnamefont{Pailh\`es}},
  \bibinfo{author}{\bibfnamefont{N.~S.} \bibnamefont{Berzigiarova}},
  \bibinfo{author}{\bibfnamefont{N.~N.} \bibnamefont{Kolesnikov}},
  \bibnamefont{and} \bibinfo{author}{\bibfnamefont{B.}~\bibnamefont{Keimer}},
  \bibinfo{journal}{Science} \textbf{\bibinfo{volume}{295}},
  \bibinfo{pages}{1045} (\bibinfo{year}{2002}).

\bibitem[{\citenamefont{Kastner et~al.}(1998)\citenamefont{Kastner, Birgeneau,
  Shirane, and Endoh}}]{kast98}
\bibinfo{author}{\bibfnamefont{M.~A.} \bibnamefont{Kastner}},
  \bibinfo{author}{\bibfnamefont{R.~J.} \bibnamefont{Birgeneau}},
  \bibinfo{author}{\bibfnamefont{G.}~\bibnamefont{Shirane}}, \bibnamefont{and}
  \bibinfo{author}{\bibfnamefont{Y.}~\bibnamefont{Endoh}},
  \bibinfo{journal}{Rev. Mod. Phys.} \textbf{\bibinfo{volume}{70}},
  \bibinfo{pages}{897} (\bibinfo{year}{1998}).

\bibitem[{\citenamefont{Schachinger and Carbotte}(2000)}]{scha00}
\bibinfo{author}{\bibfnamefont{E.}~\bibnamefont{Schachinger}} \bibnamefont{and}
  \bibinfo{author}{\bibfnamefont{J.~P.} \bibnamefont{Carbotte}},
  \bibinfo{journal}{Phys. Rev. B} \textbf{\bibinfo{volume}{62}},
  \bibinfo{pages}{9054} (\bibinfo{year}{2000}).

\bibitem[{\citenamefont{Eschrig and Norman}(2000)}]{esch00}
\bibinfo{author}{\bibfnamefont{M.}~\bibnamefont{Eschrig}} \bibnamefont{and}
  \bibinfo{author}{\bibfnamefont{M.~R.} \bibnamefont{Norman}},
  \bibinfo{journal}{Phys. Rev. Lett.} \textbf{\bibinfo{volume}{85}},
  \bibinfo{pages}{3261} (\bibinfo{year}{2000}).

\bibitem[{\citenamefont{Singley et~al.}(2001)\citenamefont{Singley, Basov,
  Kurahashi, Uefuji, and Yamada}}]{sing01}
\bibinfo{author}{\bibfnamefont{E.~J.} \bibnamefont{Singley}},
  \bibinfo{author}{\bibfnamefont{D.~N.} \bibnamefont{Basov}},
  \bibinfo{author}{\bibfnamefont{K.}~\bibnamefont{Kurahashi}},
  \bibinfo{author}{\bibfnamefont{T.}~\bibnamefont{Uefuji}}, \bibnamefont{and}
  \bibinfo{author}{\bibfnamefont{K.}~\bibnamefont{Yamada}},
  \bibinfo{journal}{Phys. Rev. B} \textbf{\bibinfo{volume}{64}},
  \bibinfo{pages}{224503} (\bibinfo{year}{2001}).

\bibitem[{\citenamefont{Zhou et~al.}(2003)\citenamefont{Zhou, Yoshida, Lanzara,
  Bogdanov, Kellar, Shen, Yang, Ronning, Sasagawa, Kakeshita et~al.}}]{zhou03}
\bibinfo{author}{\bibfnamefont{X.~J.} \bibnamefont{Zhou}},
  \bibinfo{author}{\bibfnamefont{T.}~\bibnamefont{Yoshida}},
  \bibinfo{author}{\bibfnamefont{A.}~\bibnamefont{Lanzara}},
  \bibinfo{author}{\bibfnamefont{P.~V.} \bibnamefont{Bogdanov}},
  \bibinfo{author}{\bibfnamefont{S.~A.} \bibnamefont{Kellar}},
  \bibinfo{author}{\bibfnamefont{K.~M.} \bibnamefont{Shen}},
  \bibinfo{author}{\bibfnamefont{W.~L.} \bibnamefont{Yang}},
  \bibinfo{author}{\bibfnamefont{F.}~\bibnamefont{Ronning}},
  \bibinfo{author}{\bibfnamefont{T.}~\bibnamefont{Sasagawa}},
  \bibinfo{author}{\bibfnamefont{T.}~\bibnamefont{Kakeshita}},
  \bibnamefont{et~al.}, \bibinfo{journal}{Nature}
  \textbf{\bibinfo{volume}{423}}, \bibinfo{pages}{398} (\bibinfo{year}{2003}).

\bibitem[{\citenamefont{Mason et~al.}(1996)\citenamefont{Mason, Schr\"oder,
  Aeppli, Mook, and Hayden}}]{maso96}
\bibinfo{author}{\bibfnamefont{T.~E.} \bibnamefont{Mason}},
  \bibinfo{author}{\bibfnamefont{A.}~\bibnamefont{Schr\"oder}},
  \bibinfo{author}{\bibfnamefont{G.}~\bibnamefont{Aeppli}},
  \bibinfo{author}{\bibfnamefont{H.~A.} \bibnamefont{Mook}}, \bibnamefont{and}
  \bibinfo{author}{\bibfnamefont{S.~M.} \bibnamefont{Hayden}},
  \bibinfo{journal}{Phys. Rev. Lett.} \textbf{\bibinfo{volume}{77}},
  \bibinfo{pages}{1604} (\bibinfo{year}{1996}).

\bibitem[{\citenamefont{Lake et~al.}(1999)\citenamefont{Lake, Aeppli, Mason,
  Schr\"oder, McMorrow, Lefmann, Isshiki, Nohara, Takagi, and Hayden}}]{lake99}
\bibinfo{author}{\bibfnamefont{B.}~\bibnamefont{Lake}},
  \bibinfo{author}{\bibfnamefont{G.}~\bibnamefont{Aeppli}},
  \bibinfo{author}{\bibfnamefont{T.~E.} \bibnamefont{Mason}},
  \bibinfo{author}{\bibfnamefont{A.}~\bibnamefont{Schr\"oder}},
  \bibinfo{author}{\bibfnamefont{D.~F.} \bibnamefont{McMorrow}},
  \bibinfo{author}{\bibfnamefont{K.}~\bibnamefont{Lefmann}},
  \bibinfo{author}{\bibfnamefont{M.}~\bibnamefont{Isshiki}},
  \bibinfo{author}{\bibfnamefont{M.}~\bibnamefont{Nohara}},
  \bibinfo{author}{\bibfnamefont{H.}~\bibnamefont{Takagi}}, \bibnamefont{and}
  \bibinfo{author}{\bibfnamefont{S.~M.} \bibnamefont{Hayden}},
  \bibinfo{journal}{Nature} \textbf{\bibinfo{volume}{400}}, \bibinfo{pages}{43}
  (\bibinfo{year}{1999}).

\bibitem[{\citenamefont{Dai et~al.}(2000)\citenamefont{Dai, Mook, Aeppli,
  Hayden, and Do\u{g}an}}]{dai00}
\bibinfo{author}{\bibfnamefont{P.}~\bibnamefont{Dai}},
  \bibinfo{author}{\bibfnamefont{H.~A.} \bibnamefont{Mook}},
  \bibinfo{author}{\bibfnamefont{G.}~\bibnamefont{Aeppli}},
  \bibinfo{author}{\bibfnamefont{S.~M.} \bibnamefont{Hayden}},
  \bibnamefont{and}
  \bibinfo{author}{\bibfnamefont{F.}~\bibnamefont{Do\u{g}an}},
  \bibinfo{journal}{Nature} \textbf{\bibinfo{volume}{406}},
  \bibinfo{pages}{965} (\bibinfo{year}{2000}).

\bibitem[{\citenamefont{Katano et~al.}(2000)\citenamefont{Katano, Sato, Yamada,
  Suzuki, and Fukase}}]{kata00}
\bibinfo{author}{\bibfnamefont{S.}~\bibnamefont{Katano}},
  \bibinfo{author}{\bibfnamefont{M.}~\bibnamefont{Sato}},
  \bibinfo{author}{\bibfnamefont{K.}~\bibnamefont{Yamada}},
  \bibinfo{author}{\bibfnamefont{T.}~\bibnamefont{Suzuki}}, \bibnamefont{and}
  \bibinfo{author}{\bibfnamefont{T.}~\bibnamefont{Fukase}},
  \bibinfo{journal}{Phys. Rev. B} \textbf{\bibinfo{volume}{62}},
  \bibinfo{pages}{R14677} (\bibinfo{year}{2000}).

\bibitem[{\citenamefont{Lake et~al.}(2002)\citenamefont{Lake, {R\o nnow},
  Christensen, Aeppli, Lefmann, McMorrow, Vorderwisch, Smeibidl, Mangkorntong,
  Sasagawa et~al.}}]{lake02}
\bibinfo{author}{\bibfnamefont{B.}~\bibnamefont{Lake}},
  \bibinfo{author}{\bibfnamefont{H.~M.} \bibnamefont{{R\o nnow}}},
  \bibinfo{author}{\bibfnamefont{N.~B.} \bibnamefont{Christensen}},
  \bibinfo{author}{\bibfnamefont{G.}~\bibnamefont{Aeppli}},
  \bibinfo{author}{\bibfnamefont{K.}~\bibnamefont{Lefmann}},
  \bibinfo{author}{\bibfnamefont{D.~F.} \bibnamefont{McMorrow}},
  \bibinfo{author}{\bibfnamefont{P.}~\bibnamefont{Vorderwisch}},
  \bibinfo{author}{\bibfnamefont{P.}~\bibnamefont{Smeibidl}},
  \bibinfo{author}{\bibfnamefont{N.}~\bibnamefont{Mangkorntong}},
  \bibinfo{author}{\bibfnamefont{T.}~\bibnamefont{Sasagawa}},
  \bibnamefont{et~al.}, \bibinfo{journal}{Nature}
  \textbf{\bibinfo{volume}{415}}, \bibinfo{pages}{299} (\bibinfo{year}{2002}).

\bibitem[{\citenamefont{Khaykovich et~al.}(2002)\citenamefont{Khaykovich, Lee,
  Erwin, Lee, Wakimoto, Thomas, Kastner, and Birgeneau}}]{khay02}
\bibinfo{author}{\bibfnamefont{B.}~\bibnamefont{Khaykovich}},
  \bibinfo{author}{\bibfnamefont{Y.~S.} \bibnamefont{Lee}},
  \bibinfo{author}{\bibfnamefont{R.~W.} \bibnamefont{Erwin}},
  \bibinfo{author}{\bibfnamefont{S.-H.} \bibnamefont{Lee}},
  \bibinfo{author}{\bibfnamefont{S.}~\bibnamefont{Wakimoto}},
  \bibinfo{author}{\bibfnamefont{K.~J.} \bibnamefont{Thomas}},
  \bibinfo{author}{\bibfnamefont{M.~A.} \bibnamefont{Kastner}},
  \bibnamefont{and} \bibinfo{author}{\bibfnamefont{R.~J.}
  \bibnamefont{Birgeneau}}, \bibinfo{journal}{Phys. Rev. B}
  \textbf{\bibinfo{volume}{66}}, \bibinfo{pages}{014528}
  (\bibinfo{year}{2002}).

\bibitem[{\citenamefont{Khaykovich et~al.}(2003)\citenamefont{Khaykovich,
  Birgeneau, Chou, Erwin, Kastner, Lee, Lee, Smeibidl, Vorderwisch, and
  Wakimoto}}]{khay03}
\bibinfo{author}{\bibfnamefont{B.}~\bibnamefont{Khaykovich}},
  \bibinfo{author}{\bibfnamefont{R.~J.} \bibnamefont{Birgeneau}},
  \bibinfo{author}{\bibfnamefont{F.~C.} \bibnamefont{Chou}},
  \bibinfo{author}{\bibfnamefont{R.~W.} \bibnamefont{Erwin}},
  \bibinfo{author}{\bibfnamefont{M.~A.} \bibnamefont{Kastner}},
  \bibinfo{author}{\bibfnamefont{S.-H.} \bibnamefont{Lee}},
  \bibinfo{author}{\bibfnamefont{Y.~S.} \bibnamefont{Lee}},
  \bibinfo{author}{\bibfnamefont{P.}~\bibnamefont{Smeibidl}},
  \bibinfo{author}{\bibfnamefont{P.}~\bibnamefont{Vorderwisch}},
  \bibnamefont{and} \bibinfo{author}{\bibfnamefont{S.}~\bibnamefont{Wakimoto}},
  \bibinfo{journal}{Phys. Rev. B} \textbf{\bibinfo{volume}{67}},
  \bibinfo{pages}{054501} (\bibinfo{year}{2003}).

\bibitem[{\citenamefont{Lake et~al.}(2001)\citenamefont{Lake, Aeppli, Clausen,
  McMorrow, Lefmann, Hussey, Mangkorntong, Nohara, Takagi, Mason
  et~al.}}]{lake01}
\bibinfo{author}{\bibfnamefont{B.}~\bibnamefont{Lake}},
  \bibinfo{author}{\bibfnamefont{G.}~\bibnamefont{Aeppli}},
  \bibinfo{author}{\bibfnamefont{K.~N.} \bibnamefont{Clausen}},
  \bibinfo{author}{\bibfnamefont{D.~F.} \bibnamefont{McMorrow}},
  \bibinfo{author}{\bibfnamefont{K.}~\bibnamefont{Lefmann}},
  \bibinfo{author}{\bibfnamefont{N.~E.} \bibnamefont{Hussey}},
  \bibinfo{author}{\bibfnamefont{N.}~\bibnamefont{Mangkorntong}},
  \bibinfo{author}{\bibfnamefont{M.}~\bibnamefont{Nohara}},
  \bibinfo{author}{\bibfnamefont{H.}~\bibnamefont{Takagi}},
  \bibinfo{author}{\bibfnamefont{T.~E.} \bibnamefont{Mason}},
  \bibnamefont{et~al.}, \bibinfo{journal}{Science}
  \textbf{\bibinfo{volume}{291}}, \bibinfo{pages}{1759} (\bibinfo{year}{2001}).

\bibitem[{\citenamefont{Kimura et~al.}(2003)\citenamefont{Kimura, Kofu,
  Matsumoto, and Hirota}}]{kimu03b}
\bibinfo{author}{\bibfnamefont{H.}~\bibnamefont{Kimura}},
  \bibinfo{author}{\bibfnamefont{M.}~\bibnamefont{Kofu}},
  \bibinfo{author}{\bibfnamefont{Y.}~\bibnamefont{Matsumoto}},
  \bibnamefont{and} \bibinfo{author}{\bibfnamefont{K.}~\bibnamefont{Hirota}},
  \bibinfo{journal}{Phys. Rev. Lett.} \textbf{\bibinfo{volume}{91}},
  \bibinfo{pages}{067002} (\bibinfo{year}{2003}).

\bibitem[{\citenamefont{Lee et~al.}(2003)\citenamefont{Lee, Yamada, Hiraka,
  {Venkateswara Rao}, and Endoh}}]{lee03}
\bibinfo{author}{\bibfnamefont{C.~H.} \bibnamefont{Lee}},
  \bibinfo{author}{\bibfnamefont{K.}~\bibnamefont{Yamada}},
  \bibinfo{author}{\bibfnamefont{H.}~\bibnamefont{Hiraka}},
  \bibinfo{author}{\bibfnamefont{C.~R.} \bibnamefont{{Venkateswara Rao}}},
  \bibnamefont{and} \bibinfo{author}{\bibfnamefont{Y.}~\bibnamefont{Endoh}},
  \bibinfo{journal}{Phys. Rev. B} \textbf{\bibinfo{volume}{67}},
  \bibinfo{pages}{134521} (\bibinfo{year}{2003}).

\bibitem[{\citenamefont{Shirane et~al.}(2002)\citenamefont{Shirane, Shapiro,
  and Tranquada}}]{shir02}
\bibinfo{author}{\bibfnamefont{G.}~\bibnamefont{Shirane}},
  \bibinfo{author}{\bibfnamefont{S.~M.} \bibnamefont{Shapiro}},
  \bibnamefont{and} \bibinfo{author}{\bibfnamefont{J.~M.}
  \bibnamefont{Tranquada}}, \emph{\bibinfo{title}{Neutron Scattering with a
  Triple-Axis Spectrometer: Basic Techniques}} (\bibinfo{publisher}{Cambridge
  University Press}, \bibinfo{address}{Cambridge}, \bibinfo{year}{2002}).

\bibitem[{\citenamefont{Li et~al.}(1993)\citenamefont{Li, Suenaga, Kimura, and
  Kishio}}]{li93}
\bibinfo{author}{\bibfnamefont{Q.}~\bibnamefont{Li}},
  \bibinfo{author}{\bibfnamefont{M.}~\bibnamefont{Suenaga}},
  \bibinfo{author}{\bibfnamefont{T.}~\bibnamefont{Kimura}}, \bibnamefont{and}
  \bibinfo{author}{\bibfnamefont{K.}~\bibnamefont{Kishio}},
  \bibinfo{journal}{Phys. Rev. B} \textbf{\bibinfo{volume}{47}},
  \bibinfo{pages}{11384} (\bibinfo{year}{1993}).

\bibitem[{\citenamefont{Momono et~al.}(1998)\citenamefont{Momono, Nakano, Oda,
  and Ido}}]{momo98}
\bibinfo{author}{\bibfnamefont{N.}~\bibnamefont{Momono}},
  \bibinfo{author}{\bibfnamefont{T.}~\bibnamefont{Nakano}},
  \bibinfo{author}{\bibfnamefont{M.}~\bibnamefont{Oda}}, \bibnamefont{and}
  \bibinfo{author}{\bibfnamefont{M.}~\bibnamefont{Ido}}, \bibinfo{journal}{J.
  Phys. Chem. Solids} \textbf{\bibinfo{volume}{59}}, \bibinfo{pages}{2068}
  (\bibinfo{year}{1998}).

\bibitem[{\citenamefont{Chen et~al.}(1994)\citenamefont{Chen, Irwin, Trodahl,
  Kimura, and Kishio}}]{chen94}
\bibinfo{author}{\bibfnamefont{X.~K.} \bibnamefont{Chen}},
  \bibinfo{author}{\bibfnamefont{J.~C.} \bibnamefont{Irwin}},
  \bibinfo{author}{\bibfnamefont{H.~J.} \bibnamefont{Trodahl}},
  \bibinfo{author}{\bibfnamefont{T.}~\bibnamefont{Kimura}}, \bibnamefont{and}
  \bibinfo{author}{\bibfnamefont{K.}~\bibnamefont{Kishio}},
  \bibinfo{journal}{Phys. Rev. Lett.} \textbf{\bibinfo{volume}{73}},
  \bibinfo{pages}{3290} (\bibinfo{year}{1994}).

\bibitem[{\citenamefont{Rossat-Mignod et~al.}(1991)\citenamefont{Rossat-Mignod,
  Regnault, Vettier, Burlet, Henry, and Lapertot}}]{ross91}
\bibinfo{author}{\bibfnamefont{J.}~\bibnamefont{Rossat-Mignod}},
  \bibinfo{author}{\bibfnamefont{L.~P.} \bibnamefont{Regnault}},
  \bibinfo{author}{\bibfnamefont{C.}~\bibnamefont{Vettier}},
  \bibinfo{author}{\bibfnamefont{P.}~\bibnamefont{Burlet}},
  \bibinfo{author}{\bibfnamefont{J.~Y.} \bibnamefont{Henry}}, \bibnamefont{and}
  \bibinfo{author}{\bibfnamefont{G.}~\bibnamefont{Lapertot}},
  \bibinfo{journal}{Physica B} \textbf{\bibinfo{volume}{169}},
  \bibinfo{pages}{58} (\bibinfo{year}{1991}).

\bibitem[{\citenamefont{Bourges et~al.}(2000)\citenamefont{Bourges, Sidis,
  Fong, Regnault, Bossy, Ivanov, and Keimer}}]{bour00}
\bibinfo{author}{\bibfnamefont{P.}~\bibnamefont{Bourges}},
  \bibinfo{author}{\bibfnamefont{Y.}~\bibnamefont{Sidis}},
  \bibinfo{author}{\bibfnamefont{H.~F.} \bibnamefont{Fong}},
  \bibinfo{author}{\bibfnamefont{L.~P.} \bibnamefont{Regnault}},
  \bibinfo{author}{\bibfnamefont{J.}~\bibnamefont{Bossy}},
  \bibinfo{author}{\bibfnamefont{A.}~\bibnamefont{Ivanov}}, \bibnamefont{and}
  \bibinfo{author}{\bibfnamefont{B.}~\bibnamefont{Keimer}},
  \bibinfo{journal}{Science} \textbf{\bibinfo{volume}{288}},
  \bibinfo{pages}{1234} (\bibinfo{year}{2000}).

\bibitem[{\citenamefont{Dai et~al.}(1998)\citenamefont{Dai, Mook, and {Do\u
  gan}}}]{dai98}
\bibinfo{author}{\bibfnamefont{P.}~\bibnamefont{Dai}},
  \bibinfo{author}{\bibfnamefont{H.~A.} \bibnamefont{Mook}}, \bibnamefont{and}
  \bibinfo{author}{\bibfnamefont{F.}~\bibnamefont{{Do\u gan}}},
  \bibinfo{journal}{Phys. Rev. Lett.} \textbf{\bibinfo{volume}{80}},
  \bibinfo{pages}{1738} (\bibinfo{year}{1998}).

\bibitem[{\citenamefont{Reznik et~al.}(2003)\citenamefont{Reznik, Bourges,
  Pintschovius, Endoh, Sidis, Shiokara, and Tajima}}]{rezn03}
\bibinfo{author}{\bibfnamefont{D.}~\bibnamefont{Reznik}},
  \bibinfo{author}{\bibfnamefont{P.}~\bibnamefont{Bourges}},
  \bibinfo{author}{\bibfnamefont{L.}~\bibnamefont{Pintschovius}},
  \bibinfo{author}{\bibfnamefont{Y.}~\bibnamefont{Endoh}},
  \bibinfo{author}{\bibfnamefont{Y.}~\bibnamefont{Sidis}},
  \bibinfo{author}{\bibfnamefont{Y.}~\bibnamefont{Shiokara}}, \bibnamefont{and}
  \bibinfo{author}{\bibfnamefont{S.}~\bibnamefont{Tajima}}
  (\bibinfo{year}{2003}), \eprint{cond-mat/0307591}.

\bibitem[{\citenamefont{Demler and Zhang}(1995)}]{deml95}
\bibinfo{author}{\bibfnamefont{E.}~\bibnamefont{Demler}} \bibnamefont{and}
  \bibinfo{author}{\bibfnamefont{S.-C.} \bibnamefont{Zhang}},
  \bibinfo{journal}{Phys. Rev. Lett.} \textbf{\bibinfo{volume}{75}},
  \bibinfo{pages}{4126} (\bibinfo{year}{1995}).

\bibitem[{\citenamefont{Zhang}(1997)}]{zhan97}
\bibinfo{author}{\bibfnamefont{S.-C.} \bibnamefont{Zhang}},
  \bibinfo{journal}{Science} \textbf{\bibinfo{volume}{275}},
  \bibinfo{pages}{1089} (\bibinfo{year}{1997}).

\bibitem[{\citenamefont{Veillette et~al.}(1999)\citenamefont{Veillette,
  Bazaliy, Berlinsky, and Kallin}}]{veil99}
\bibinfo{author}{\bibfnamefont{M.}~\bibnamefont{Veillette}},
  \bibinfo{author}{\bibfnamefont{Y.~B.} \bibnamefont{Bazaliy}},
  \bibinfo{author}{\bibfnamefont{A.~J.} \bibnamefont{Berlinsky}},
  \bibnamefont{and} \bibinfo{author}{\bibfnamefont{C.}~\bibnamefont{Kallin}},
  \bibinfo{journal}{Phys. Rev. Lett.} \textbf{\bibinfo{volume}{83}},
  \bibinfo{pages}{2413} (\bibinfo{year}{1999}).

\bibitem[{\citenamefont{Hu and Zhang}(2001)}]{hu01}
\bibinfo{author}{\bibfnamefont{J.-P.} \bibnamefont{Hu}} \bibnamefont{and}
  \bibinfo{author}{\bibfnamefont{S.-C.} \bibnamefont{Zhang}},
  \bibinfo{journal}{Phys. Rev. B} \textbf{\bibinfo{volume}{64}},
  \bibinfo{pages}{100502(R)} (\bibinfo{year}{2001}).

\bibitem[{\citenamefont{Morr and Pines}(1998)}]{morr98}
\bibinfo{author}{\bibfnamefont{D.~K.} \bibnamefont{Morr}} \bibnamefont{and}
  \bibinfo{author}{\bibfnamefont{D.}~\bibnamefont{Pines}},
  \bibinfo{journal}{Phys. Rev. Lett.} \textbf{\bibinfo{volume}{81}},
  \bibinfo{pages}{1086} (\bibinfo{year}{1998}).

\bibitem[{\citenamefont{Abanov and Chubukov}(1999)}]{aban99}
\bibinfo{author}{\bibfnamefont{A.}~\bibnamefont{Abanov}} \bibnamefont{and}
  \bibinfo{author}{\bibfnamefont{A.~V.} \bibnamefont{Chubukov}},
  \bibinfo{journal}{Phys. Rev. Lett.} \textbf{\bibinfo{volume}{83}},
  \bibinfo{pages}{1652} (\bibinfo{year}{1999}).

\bibitem[{\citenamefont{Morr and Pines}(2000)}]{morr00}
\bibinfo{author}{\bibfnamefont{D.~K.} \bibnamefont{Morr}} \bibnamefont{and}
  \bibinfo{author}{\bibfnamefont{D.}~\bibnamefont{Pines}},
  \bibinfo{journal}{Phys. Rev. B} \textbf{\bibinfo{volume}{61}},
  \bibinfo{pages}{R6483} (\bibinfo{year}{2000}).

\bibitem[{\citenamefont{Lavagna and Stemmann}(1994)}]{lava94}
\bibinfo{author}{\bibfnamefont{M.}~\bibnamefont{Lavagna}} \bibnamefont{and}
  \bibinfo{author}{\bibfnamefont{G.}~\bibnamefont{Stemmann}},
  \bibinfo{journal}{Phys. Rev. B} \textbf{\bibinfo{volume}{49}},
  \bibinfo{pages}{4235} (\bibinfo{year}{1994}).

\bibitem[{\citenamefont{Mazin and Yakovenko}(1995)}]{mazi95}
\bibinfo{author}{\bibfnamefont{I.~I.} \bibnamefont{Mazin}} \bibnamefont{and}
  \bibinfo{author}{\bibfnamefont{V.~M.} \bibnamefont{Yakovenko}},
  \bibinfo{journal}{Phys. Rev. Lett.} \textbf{\bibinfo{volume}{75}},
  \bibinfo{pages}{4134} (\bibinfo{year}{1995}).

\bibitem[{\citenamefont{Bulut and Scalapino}(1996)}]{bulu96}
\bibinfo{author}{\bibfnamefont{N.}~\bibnamefont{Bulut}} \bibnamefont{and}
  \bibinfo{author}{\bibfnamefont{D.~J.} \bibnamefont{Scalapino}},
  \bibinfo{journal}{Phys. Rev. B} \textbf{\bibinfo{volume}{53}},
  \bibinfo{pages}{5149} (\bibinfo{year}{1996}).

\bibitem[{\citenamefont{Dahm et~al.}(1998)\citenamefont{Dahm, Manske, and
  Tewordt}}]{dahm98}
\bibinfo{author}{\bibfnamefont{T.}~\bibnamefont{Dahm}},
  \bibinfo{author}{\bibfnamefont{D.}~\bibnamefont{Manske}}, \bibnamefont{and}
  \bibinfo{author}{\bibfnamefont{L.}~\bibnamefont{Tewordt}},
  \bibinfo{journal}{Phys. Rev. B} \textbf{\bibinfo{volume}{58}},
  \bibinfo{pages}{12454} (\bibinfo{year}{1998}).

\bibitem[{\citenamefont{Brinckmann and Lee}(1999)}]{brin99}
\bibinfo{author}{\bibfnamefont{J.}~\bibnamefont{Brinckmann}} \bibnamefont{and}
  \bibinfo{author}{\bibfnamefont{P.~A.} \bibnamefont{Lee}},
  \bibinfo{journal}{Phys. Rev. Lett.} \textbf{\bibinfo{volume}{82}},
  \bibinfo{pages}{2915} (\bibinfo{year}{1999}).

\bibitem[{\citenamefont{Kao et~al.}(2000)\citenamefont{Kao, Si, and
  Levin}}]{kao00}
\bibinfo{author}{\bibfnamefont{Y.-J.} \bibnamefont{Kao}},
  \bibinfo{author}{\bibfnamefont{Q.}~\bibnamefont{Si}}, \bibnamefont{and}
  \bibinfo{author}{\bibfnamefont{K.}~\bibnamefont{Levin}},
  \bibinfo{journal}{Phys. Rev. B} \textbf{\bibinfo{volume}{61}},
  \bibinfo{pages}{R11898} (\bibinfo{year}{2000}).

\bibitem[{\citenamefont{Onufrieva and Pfeuty}(2002)}]{onuf02}
\bibinfo{author}{\bibfnamefont{F.}~\bibnamefont{Onufrieva}} \bibnamefont{and}
  \bibinfo{author}{\bibfnamefont{P.}~\bibnamefont{Pfeuty}},
  \bibinfo{journal}{Phys. Rev. B} \textbf{\bibinfo{volume}{65}},
  \bibinfo{pages}{054515} (\bibinfo{year}{2002}).

\bibitem[{\citenamefont{Norman}(2000)}]{norm00}
\bibinfo{author}{\bibfnamefont{M.~R.} \bibnamefont{Norman}},
  \bibinfo{journal}{Phys. Rev. B} \textbf{\bibinfo{volume}{61}},
  \bibinfo{pages}{14751} (\bibinfo{year}{2000}).

\bibitem[{\citenamefont{Damascelli et~al.}(2003)\citenamefont{Damascelli, Shen,
  and Hussain}}]{dama03}
\bibinfo{author}{\bibfnamefont{A.}~\bibnamefont{Damascelli}},
  \bibinfo{author}{\bibfnamefont{Z.-X.} \bibnamefont{Shen}}, \bibnamefont{and}
  \bibinfo{author}{\bibfnamefont{Z.}~\bibnamefont{Hussain}},
  \bibinfo{journal}{Rev. Mod. Phys.} \textbf{\bibinfo{volume}{75}},
  \bibinfo{pages}{473} (\bibinfo{year}{2003}).

\bibitem[{\citenamefont{Borisenko et~al.}(2003)\citenamefont{Borisenko,
  Kordyuk, Kim, Koitzsch, Knupfer, Golden, Fink, Eschrig, Berger, and
  Follath}}]{bori03}
\bibinfo{author}{\bibfnamefont{S.~V.} \bibnamefont{Borisenko}},
  \bibinfo{author}{\bibfnamefont{A.~A.} \bibnamefont{Kordyuk}},
  \bibinfo{author}{\bibfnamefont{T.~K.} \bibnamefont{Kim}},
  \bibinfo{author}{\bibfnamefont{A.}~\bibnamefont{Koitzsch}},
  \bibinfo{author}{\bibfnamefont{M.}~\bibnamefont{Knupfer}},
  \bibinfo{author}{\bibfnamefont{M.~S.} \bibnamefont{Golden}},
  \bibinfo{author}{\bibfnamefont{J.}~\bibnamefont{Fink}},
  \bibinfo{author}{\bibfnamefont{M.}~\bibnamefont{Eschrig}},
  \bibinfo{author}{\bibfnamefont{H.}~\bibnamefont{Berger}}, \bibnamefont{and}
  \bibinfo{author}{\bibfnamefont{R.}~\bibnamefont{Follath}},
  \bibinfo{journal}{Phys. Rev. Lett.} \textbf{\bibinfo{volume}{90}},
  \bibinfo{pages}{207001} (\bibinfo{year}{2003}).

\bibitem[{\citenamefont{Johnson et~al.}(2001)\citenamefont{Johnson, Valla,
  Federov, Yusof, Wells, Li, Moodenbaugh, Gu, Koshizuka, Kendziora
  et~al.}}]{john01}
\bibinfo{author}{\bibfnamefont{P.~D.} \bibnamefont{Johnson}},
  \bibinfo{author}{\bibfnamefont{T.}~\bibnamefont{Valla}},
  \bibinfo{author}{\bibfnamefont{A.~V.} \bibnamefont{Federov}},
  \bibinfo{author}{\bibfnamefont{Z.}~\bibnamefont{Yusof}},
  \bibinfo{author}{\bibfnamefont{B.~O.} \bibnamefont{Wells}},
  \bibinfo{author}{\bibfnamefont{Q.}~\bibnamefont{Li}},
  \bibinfo{author}{\bibfnamefont{A.~R.} \bibnamefont{Moodenbaugh}},
  \bibinfo{author}{\bibfnamefont{G.~D.} \bibnamefont{Gu}},
  \bibinfo{author}{\bibfnamefont{N.}~\bibnamefont{Koshizuka}},
  \bibinfo{author}{\bibfnamefont{C.}~\bibnamefont{Kendziora}},
  \bibnamefont{et~al.}, \bibinfo{journal}{Phys. Rev. Lett.}
  \textbf{\bibinfo{volume}{87}}, \bibinfo{pages}{177007}
  (\bibinfo{year}{2001}).

\bibitem[{\citenamefont{Kee et~al.}(2002)\citenamefont{Kee, Kivelson, and
  Aeppli}}]{kee02}
\bibinfo{author}{\bibfnamefont{H.-Y.} \bibnamefont{Kee}},
  \bibinfo{author}{\bibfnamefont{S.~A.} \bibnamefont{Kivelson}},
  \bibnamefont{and} \bibinfo{author}{\bibfnamefont{G.}~\bibnamefont{Aeppli}},
  \bibinfo{journal}{Phys. Rev. Lett.} \textbf{\bibinfo{volume}{88}},
  \bibinfo{pages}{257002} (\bibinfo{year}{2002}).

\bibitem[{\citenamefont{Tranquada et~al.}(1999)\citenamefont{Tranquada,
  Ichikawa, and Uchida}}]{tran99a}
\bibinfo{author}{\bibfnamefont{J.~M.} \bibnamefont{Tranquada}},
  \bibinfo{author}{\bibfnamefont{N.}~\bibnamefont{Ichikawa}}, \bibnamefont{and}
  \bibinfo{author}{\bibfnamefont{S.}~\bibnamefont{Uchida}},
  \bibinfo{journal}{Phys. Rev. B} \textbf{\bibinfo{volume}{59}},
  \bibinfo{pages}{14712} (\bibinfo{year}{1999}).

\bibitem[{\citenamefont{Ito et~al.}(2003)\citenamefont{Ito, Yasui, Iikubo,
  Sato, Sato, Kobayashi, and Kakurai}}]{ito03}
\bibinfo{author}{\bibfnamefont{M.}~\bibnamefont{Ito}},
  \bibinfo{author}{\bibfnamefont{Y.}~\bibnamefont{Yasui}},
  \bibinfo{author}{\bibfnamefont{S.}~\bibnamefont{Iikubo}},
  \bibinfo{author}{\bibfnamefont{M.}~\bibnamefont{Sato}},
  \bibinfo{author}{\bibfnamefont{M.}~\bibnamefont{Sato}},
  \bibinfo{author}{\bibfnamefont{A.}~\bibnamefont{Kobayashi}},
  \bibnamefont{and} \bibinfo{author}{\bibfnamefont{K.}~\bibnamefont{Kakurai}},
  \bibinfo{journal}{J. Phys. Soc. Japan} \textbf{\bibinfo{volume}{72}},
  \bibinfo{pages}{1627} (\bibinfo{year}{2003}).

\bibitem[{\citenamefont{Chubukov et~al.}(1994)\citenamefont{Chubukov, Sachdev,
  and Ye}}]{chub94}
\bibinfo{author}{\bibfnamefont{A.~V.} \bibnamefont{Chubukov}},
  \bibinfo{author}{\bibfnamefont{S.}~\bibnamefont{Sachdev}}, \bibnamefont{and}
  \bibinfo{author}{\bibfnamefont{J.}~\bibnamefont{Ye}}, \bibinfo{journal}{Phys.
  Rev. B} \textbf{\bibinfo{volume}{49}}, \bibinfo{pages}{11919}
  (\bibinfo{year}{1994}).

\bibitem[{\citenamefont{Kivelson et~al.}()\citenamefont{Kivelson, Bindloss,
  Fradkin, Oganesyan, Tranquada, Kapitulnik, and Howald}}]{kive03}
\bibinfo{author}{\bibfnamefont{S.~A.} \bibnamefont{Kivelson}},
  \bibinfo{author}{\bibfnamefont{I.~P.} \bibnamefont{Bindloss}},
  \bibinfo{author}{\bibfnamefont{E.}~\bibnamefont{Fradkin}},
  \bibinfo{author}{\bibfnamefont{V.}~\bibnamefont{Oganesyan}},
  \bibinfo{author}{\bibfnamefont{J.~M.} \bibnamefont{Tranquada}},
  \bibinfo{author}{\bibfnamefont{A.}~\bibnamefont{Kapitulnik}},
  \bibnamefont{and} \bibinfo{author}{\bibfnamefont{C.}~\bibnamefont{Howald}},
  \bibinfo{note}{\null Rev. Mod. Phys. (accepted); cond-mat/0210683}.

\bibitem[{\citenamefont{Zaanen and Gunnarsson}(1989)}]{zaan89}
\bibinfo{author}{\bibfnamefont{J.}~\bibnamefont{Zaanen}} \bibnamefont{and}
  \bibinfo{author}{\bibfnamefont{O.}~\bibnamefont{Gunnarsson}},
  \bibinfo{journal}{Phys. Rev. B} \textbf{\bibinfo{volume}{40}},
  \bibinfo{pages}{7391} (\bibinfo{year}{1989}).

\bibitem[{\citenamefont{Machida}(1989)}]{mach89}
\bibinfo{author}{\bibfnamefont{K.}~\bibnamefont{Machida}},
  \bibinfo{journal}{Physica C} \textbf{\bibinfo{volume}{158}},
  \bibinfo{pages}{192} (\bibinfo{year}{1989}).

\bibitem[{\citenamefont{Emery et~al.}(1999)\citenamefont{Emery, Kivelson, and
  Tranquada}}]{emer99}
\bibinfo{author}{\bibfnamefont{V.~J.} \bibnamefont{Emery}},
  \bibinfo{author}{\bibfnamefont{S.~A.} \bibnamefont{Kivelson}},
  \bibnamefont{and} \bibinfo{author}{\bibfnamefont{J.~M.}
  \bibnamefont{Tranquada}}, \bibinfo{journal}{Proc. Natl. Acad. Sci. USA}
  \textbf{\bibinfo{volume}{96}}, \bibinfo{pages}{8814} (\bibinfo{year}{1999}).

\bibitem[{\citenamefont{Emery et~al.}(1997)\citenamefont{Emery, Kivelson, and
  Zachar}}]{emer97}
\bibinfo{author}{\bibfnamefont{V.~J.} \bibnamefont{Emery}},
  \bibinfo{author}{\bibfnamefont{S.~A.} \bibnamefont{Kivelson}},
  \bibnamefont{and} \bibinfo{author}{\bibfnamefont{O.}~\bibnamefont{Zachar}},
  \bibinfo{journal}{Phys. Rev. B} \textbf{\bibinfo{volume}{56}},
  \bibinfo{pages}{6120} (\bibinfo{year}{1997}).

\bibitem[{\citenamefont{Batista et~al.}(2001)\citenamefont{Batista, Ortiz, and
  Balatsky}}]{bati01}
\bibinfo{author}{\bibfnamefont{C.~D.} \bibnamefont{Batista}},
  \bibinfo{author}{\bibfnamefont{G.}~\bibnamefont{Ortiz}}, \bibnamefont{and}
  \bibinfo{author}{\bibfnamefont{A.~V.} \bibnamefont{Balatsky}},
  \bibinfo{journal}{Phys. Rev. B} \textbf{\bibinfo{volume}{64}},
  \bibinfo{pages}{172508} (\bibinfo{year}{2001}).

\bibitem[{\citenamefont{Bourges et~al.}(2003)\citenamefont{Bourges, Sidis,
  Braden, Nakajima, and Tranquada}}]{bour03}
\bibinfo{author}{\bibfnamefont{P.}~\bibnamefont{Bourges}},
  \bibinfo{author}{\bibfnamefont{Y.}~\bibnamefont{Sidis}},
  \bibinfo{author}{\bibfnamefont{M.}~\bibnamefont{Braden}},
  \bibinfo{author}{\bibfnamefont{K.}~\bibnamefont{Nakajima}}, \bibnamefont{and}
  \bibinfo{author}{\bibfnamefont{J.~M.} \bibnamefont{Tranquada}},
  \bibinfo{journal}{Phys. Rev. Lett.} \textbf{\bibinfo{volume}{90}},
  \bibinfo{pages}{147202} (\bibinfo{year}{2003}).

\bibitem[{\citenamefont{Arovas et~al.}(1997)\citenamefont{Arovas, Berlinsky,
  Kallin, and Zhang}}]{arov97}
\bibinfo{author}{\bibfnamefont{D.~P.} \bibnamefont{Arovas}},
  \bibinfo{author}{\bibfnamefont{A.~J.} \bibnamefont{Berlinsky}},
  \bibinfo{author}{\bibfnamefont{C.}~\bibnamefont{Kallin}}, \bibnamefont{and}
  \bibinfo{author}{\bibfnamefont{S.-C.} \bibnamefont{Zhang}},
  \bibinfo{journal}{Phys. Rev. Lett.} \textbf{\bibinfo{volume}{79}},
  \bibinfo{pages}{2871} (\bibinfo{year}{1997}).

\bibitem[{\citenamefont{Franz et~al.}(2002)\citenamefont{Franz, Sheehy, and
  Te\v{s}anovi\'c}}]{fran02}
\bibinfo{author}{\bibfnamefont{M.}~\bibnamefont{Franz}},
  \bibinfo{author}{\bibfnamefont{D.~E.} \bibnamefont{Sheehy}},
  \bibnamefont{and}
  \bibinfo{author}{\bibfnamefont{Z.}~\bibnamefont{Te\v{s}anovi\'c}},
  \bibinfo{journal}{Phys. Rev. Lett.} \textbf{\bibinfo{volume}{88}},
  \bibinfo{pages}{257005} (\bibinfo{year}{2002}).

\bibitem[{\citenamefont{Chen and Ting}(2002)}]{chen02b}
\bibinfo{author}{\bibfnamefont{Y.}~\bibnamefont{Chen}} \bibnamefont{and}
  \bibinfo{author}{\bibfnamefont{C.~S.} \bibnamefont{Ting}},
  \bibinfo{journal}{Phys. Rev. B} \textbf{\bibinfo{volume}{65}},
  \bibinfo{pages}{180513} (\bibinfo{year}{2002}).

\bibitem[{\citenamefont{Lee and Wen}(2001)}]{lee01b}
\bibinfo{author}{\bibfnamefont{P.~A.} \bibnamefont{Lee}} \bibnamefont{and}
  \bibinfo{author}{\bibfnamefont{X.-G.} \bibnamefont{Wen}},
  \bibinfo{journal}{Phys. Rev. B} \textbf{\bibinfo{volume}{63}},
  \bibinfo{pages}{224517} (\bibinfo{year}{2001}).

\bibitem[{\citenamefont{Zhang et~al.}(2002)\citenamefont{Zhang, Demler, and
  Sachdev}}]{zhan02}
\bibinfo{author}{\bibfnamefont{Y.}~\bibnamefont{Zhang}},
  \bibinfo{author}{\bibfnamefont{E.}~\bibnamefont{Demler}}, \bibnamefont{and}
  \bibinfo{author}{\bibfnamefont{S.}~\bibnamefont{Sachdev}},
  \bibinfo{journal}{Phys. Rev. B} \textbf{\bibinfo{volume}{66}},
  \bibinfo{pages}{094501} (\bibinfo{year}{2002}).

\bibitem[{\citenamefont{Polkovnikov et~al.}(2002)\citenamefont{Polkovnikov,
  Vojta, and Sachdev}}]{polk02}
\bibinfo{author}{\bibfnamefont{A.}~\bibnamefont{Polkovnikov}},
  \bibinfo{author}{\bibfnamefont{M.}~\bibnamefont{Vojta}}, \bibnamefont{and}
  \bibinfo{author}{\bibfnamefont{S.}~\bibnamefont{Sachdev}},
  \bibinfo{journal}{Phys. Rev. B} \textbf{\bibinfo{volume}{65}},
  \bibinfo{pages}{220509} (\bibinfo{year}{2002}).

\bibitem[{\citenamefont{Kivelson et~al.}(2002)\citenamefont{Kivelson, Lee,
  Fradkin, and Oganesyan}}]{kive02}
\bibinfo{author}{\bibfnamefont{S.~A.} \bibnamefont{Kivelson}},
  \bibinfo{author}{\bibfnamefont{D.-H.} \bibnamefont{Lee}},
  \bibinfo{author}{\bibfnamefont{E.}~\bibnamefont{Fradkin}}, \bibnamefont{and}
  \bibinfo{author}{\bibfnamefont{V.}~\bibnamefont{Oganesyan}},
  \bibinfo{journal}{Phys. Rev. B} \textbf{\bibinfo{volume}{66}},
  \bibinfo{pages}{144516} (\bibinfo{year}{2002}).

\bibitem[{\citenamefont{Zhu et~al.}(2002)\citenamefont{Zhu, Pan, Stormer,
  Pfeiffer, and West}}]{zhu02}
\bibinfo{author}{\bibfnamefont{J.}~\bibnamefont{Zhu}},
  \bibinfo{author}{\bibfnamefont{W.}~\bibnamefont{Pan}},
  \bibinfo{author}{\bibfnamefont{H.~L.} \bibnamefont{Stormer}},
  \bibinfo{author}{\bibfnamefont{L.~N.} \bibnamefont{Pfeiffer}},
  \bibnamefont{and} \bibinfo{author}{\bibfnamefont{K.~W.} \bibnamefont{West}},
  \bibinfo{journal}{Phys. Rev. Lett.} \textbf{\bibinfo{volume}{88}},
  \bibinfo{pages}{116803} (\bibinfo{year}{2002}).

\bibitem[{\citenamefont{Chen et~al.}(2002)\citenamefont{Chen, Chen, and
  Ting}}]{chen02c}
\bibinfo{author}{\bibfnamefont{Y.}~\bibnamefont{Chen}},
  \bibinfo{author}{\bibfnamefont{H.~Y.} \bibnamefont{Chen}}, \bibnamefont{and}
  \bibinfo{author}{\bibfnamefont{C.~S.} \bibnamefont{Ting}},
  \bibinfo{journal}{Phys. Rev. B} \textbf{\bibinfo{volume}{66}},
  \bibinfo{pages}{104501} (\bibinfo{year}{2002}).

\bibitem[{\citenamefont{Takigawa et~al.}(2003)\citenamefont{Takigawa, Ichioka,
  and Machida}}]{taki03}
\bibinfo{author}{\bibfnamefont{M.}~\bibnamefont{Takigawa}},
  \bibinfo{author}{\bibfnamefont{M.}~\bibnamefont{Ichioka}}, \bibnamefont{and}
  \bibinfo{author}{\bibfnamefont{K.}~\bibnamefont{Machida}},
  \bibinfo{journal}{Phys. Rev. Lett.} \textbf{\bibinfo{volume}{90}},
  \bibinfo{pages}{047001} (\bibinfo{year}{2003}).

\bibitem[{\citenamefont{Andersen et~al.}(2003)\citenamefont{Andersen,
  Hedeg{\aa}rd, and Bruus}}]{ande03}
\bibinfo{author}{\bibfnamefont{B.~M.} \bibnamefont{Andersen}},
  \bibinfo{author}{\bibfnamefont{P.}~\bibnamefont{Hedeg{\aa}rd}},
  \bibnamefont{and} \bibinfo{author}{\bibfnamefont{H.}~\bibnamefont{Bruus}},
  \bibinfo{journal}{Phys. Rev. B} \textbf{\bibinfo{volume}{67}},
  \bibinfo{pages}{134528} (\bibinfo{year}{2003}).

\bibitem[{\citenamefont{Wakimoto et~al.}(2003)\citenamefont{Wakimoto,
  Birgeneau, Fujimaki, Ichikawa, Kasuga, Kim, Kojima, Lee, Niko, Tranquada
  et~al.}}]{waki03}
\bibinfo{author}{\bibfnamefont{S.}~\bibnamefont{Wakimoto}},
  \bibinfo{author}{\bibfnamefont{R.~J.} \bibnamefont{Birgeneau}},
  \bibinfo{author}{\bibfnamefont{Y.}~\bibnamefont{Fujimaki}},
  \bibinfo{author}{\bibfnamefont{N.}~\bibnamefont{Ichikawa}},
  \bibinfo{author}{\bibfnamefont{T.}~\bibnamefont{Kasuga}},
  \bibinfo{author}{\bibfnamefont{Y.~J.} \bibnamefont{Kim}},
  \bibinfo{author}{\bibfnamefont{K.~M.} \bibnamefont{Kojima}},
  \bibinfo{author}{\bibfnamefont{S.-H.} \bibnamefont{Lee}},
  \bibinfo{author}{\bibfnamefont{H.}~\bibnamefont{Niko}},
  \bibinfo{author}{\bibfnamefont{J.~M.} \bibnamefont{Tranquada}},
  \bibnamefont{et~al.}, \bibinfo{journal}{Phys. Rev. B}
  \textbf{\bibinfo{volume}{67}}, \bibinfo{pages}{184419}
  (\bibinfo{year}{2003}).

\bibitem[{\citenamefont{Yamada et~al.}(1995)\citenamefont{Yamada, Wakimoto,
  Shirane, Lee, Kastner, Hosoya, Greven, Endoh, and Birgeneau}}]{yama95a}
\bibinfo{author}{\bibfnamefont{K.}~\bibnamefont{Yamada}},
  \bibinfo{author}{\bibfnamefont{S.}~\bibnamefont{Wakimoto}},
  \bibinfo{author}{\bibfnamefont{G.}~\bibnamefont{Shirane}},
  \bibinfo{author}{\bibfnamefont{C.~H.} \bibnamefont{Lee}},
  \bibinfo{author}{\bibfnamefont{M.~A.} \bibnamefont{Kastner}},
  \bibinfo{author}{\bibfnamefont{S.}~\bibnamefont{Hosoya}},
  \bibinfo{author}{\bibfnamefont{M.}~\bibnamefont{Greven}},
  \bibinfo{author}{\bibfnamefont{Y.}~\bibnamefont{Endoh}}, \bibnamefont{and}
  \bibinfo{author}{\bibfnamefont{R.~J.} \bibnamefont{Birgeneau}},
  \bibinfo{journal}{Phys. Rev. Lett.} \textbf{\bibinfo{volume}{75}},
  \bibinfo{pages}{1626} (\bibinfo{year}{1995}).

\bibitem[{\citenamefont{Petit et~al.}(1997)\citenamefont{Petit, Moudden,
  Hennion, Vietkin, and Revcoleschi}}]{peti97}
\bibinfo{author}{\bibfnamefont{S.}~\bibnamefont{Petit}},
  \bibinfo{author}{\bibfnamefont{A.~H.} \bibnamefont{Moudden}},
  \bibinfo{author}{\bibfnamefont{B.}~\bibnamefont{Hennion}},
  \bibinfo{author}{\bibfnamefont{A.}~\bibnamefont{Vietkin}}, \bibnamefont{and}
  \bibinfo{author}{\bibfnamefont{A.}~\bibnamefont{Revcoleschi}},
  \bibinfo{journal}{Physica B} \textbf{\bibinfo{volume}{234--236}},
  \bibinfo{pages}{800} (\bibinfo{year}{1997}).

\bibitem[{\citenamefont{Ando et~al.}(1999)\citenamefont{Ando, Boebinger,
  Passner, Schneemeyer, Kimura, Okuya, Watauchi, Shimoyama, Kishio, Tamasaku
  et~al.}}]{ando99b}
\bibinfo{author}{\bibfnamefont{Y.}~\bibnamefont{Ando}},
  \bibinfo{author}{\bibfnamefont{G.~S.} \bibnamefont{Boebinger}},
  \bibinfo{author}{\bibfnamefont{A.}~\bibnamefont{Passner}},
  \bibinfo{author}{\bibfnamefont{L.~F.} \bibnamefont{Schneemeyer}},
  \bibinfo{author}{\bibfnamefont{T.}~\bibnamefont{Kimura}},
  \bibinfo{author}{\bibfnamefont{M.}~\bibnamefont{Okuya}},
  \bibinfo{author}{\bibfnamefont{S.}~\bibnamefont{Watauchi}},
  \bibinfo{author}{\bibfnamefont{J.}~\bibnamefont{Shimoyama}},
  \bibinfo{author}{\bibfnamefont{K.}~\bibnamefont{Kishio}},
  \bibinfo{author}{\bibfnamefont{K.}~\bibnamefont{Tamasaku}},
  \bibnamefont{et~al.}, \bibinfo{journal}{Phys. Rev. B}
  \textbf{\bibinfo{volume}{60}}, \bibinfo{pages}{12475} (\bibinfo{year}{1999}).

\bibitem[{\citenamefont{Wang et~al.}(2002)\citenamefont{Wang, Ong, Xu,
  Kakeshita, Uchida, Bonn, Liang, and Hardy}}]{wang02}
\bibinfo{author}{\bibfnamefont{Y.}~\bibnamefont{Wang}},
  \bibinfo{author}{\bibfnamefont{N.~P.} \bibnamefont{Ong}},
  \bibinfo{author}{\bibfnamefont{Z.~A.} \bibnamefont{Xu}},
  \bibinfo{author}{\bibfnamefont{T.}~\bibnamefont{Kakeshita}},
  \bibinfo{author}{\bibfnamefont{S.}~\bibnamefont{Uchida}},
  \bibinfo{author}{\bibfnamefont{D.~A.} \bibnamefont{Bonn}},
  \bibinfo{author}{\bibfnamefont{R.}~\bibnamefont{Liang}}, \bibnamefont{and}
  \bibinfo{author}{\bibfnamefont{W.~N.} \bibnamefont{Hardy}},
  \bibinfo{journal}{Phys. Rev. Lett.} \textbf{\bibinfo{volume}{88}},
  \bibinfo{pages}{257003} (\bibinfo{year}{2002}).

\bibitem[{\citenamefont{Eschrig et~al.}(2001)\citenamefont{Eschrig, Norman, and
  Jank\'o}}]{esch01}
\bibinfo{author}{\bibfnamefont{M.}~\bibnamefont{Eschrig}},
  \bibinfo{author}{\bibfnamefont{M.~R.} \bibnamefont{Norman}},
  \bibnamefont{and} \bibinfo{author}{\bibfnamefont{B.}~\bibnamefont{Jank\'o}},
  \bibinfo{journal}{Phys. Rev. B} \textbf{\bibinfo{volume}{64}},
  \bibinfo{pages}{134509} (\bibinfo{year}{2001}).

\bibitem[{\citenamefont{Hoffman et~al.}(2002)\citenamefont{Hoffman, Hudson,
  Lang, Madhavan, Eisaki, Uchida, and Davis}}]{hoff02}
\bibinfo{author}{\bibfnamefont{J.~E.} \bibnamefont{Hoffman}},
  \bibinfo{author}{\bibfnamefont{E.~W.} \bibnamefont{Hudson}},
  \bibinfo{author}{\bibfnamefont{K.~M.} \bibnamefont{Lang}},
  \bibinfo{author}{\bibfnamefont{V.}~\bibnamefont{Madhavan}},
  \bibinfo{author}{\bibfnamefont{H.}~\bibnamefont{Eisaki}},
  \bibinfo{author}{\bibfnamefont{S.}~\bibnamefont{Uchida}}, \bibnamefont{and}
  \bibinfo{author}{\bibfnamefont{J.~C.} \bibnamefont{Davis}},
  \bibinfo{journal}{Science} \textbf{\bibinfo{volume}{295}},
  \bibinfo{pages}{466} (\bibinfo{year}{2002}).

\bibitem[{\citenamefont{Nachumi et~al.}(1996)\citenamefont{Nachumi, Keren,
  Kojima, Larkin, Luke, Merrin, Tchernysh\"ov, Uemura, Ichikawa, Goto
  et~al.}}]{nach96}
\bibinfo{author}{\bibfnamefont{B.}~\bibnamefont{Nachumi}},
  \bibinfo{author}{\bibfnamefont{A.}~\bibnamefont{Keren}},
  \bibinfo{author}{\bibfnamefont{K.}~\bibnamefont{Kojima}},
  \bibinfo{author}{\bibfnamefont{M.}~\bibnamefont{Larkin}},
  \bibinfo{author}{\bibfnamefont{G.~M.} \bibnamefont{Luke}},
  \bibinfo{author}{\bibfnamefont{J.}~\bibnamefont{Merrin}},
  \bibinfo{author}{\bibfnamefont{O.}~\bibnamefont{Tchernysh\"ov}},
  \bibinfo{author}{\bibfnamefont{Y.~J.} \bibnamefont{Uemura}},
  \bibinfo{author}{\bibfnamefont{N.}~\bibnamefont{Ichikawa}},
  \bibinfo{author}{\bibfnamefont{M.}~\bibnamefont{Goto}}, \bibnamefont{et~al.},
  \bibinfo{journal}{Phys. Rev. Lett.} \textbf{\bibinfo{volume}{77}},
  \bibinfo{pages}{5421} (\bibinfo{year}{1996}).

\bibitem[{\citenamefont{Boebinger et~al.}(1996)\citenamefont{Boebinger, Ando,
  Passner, Kimura, Okuya, Shimoyama, Kishio, Tamasaku, Ichikawa, and
  Uchida}}]{boeb96}
\bibinfo{author}{\bibfnamefont{G.~S.} \bibnamefont{Boebinger}},
  \bibinfo{author}{\bibfnamefont{Y.}~\bibnamefont{Ando}},
  \bibinfo{author}{\bibfnamefont{A.}~\bibnamefont{Passner}},
  \bibinfo{author}{\bibfnamefont{T.}~\bibnamefont{Kimura}},
  \bibinfo{author}{\bibfnamefont{M.}~\bibnamefont{Okuya}},
  \bibinfo{author}{\bibfnamefont{J.}~\bibnamefont{Shimoyama}},
  \bibinfo{author}{\bibfnamefont{K.}~\bibnamefont{Kishio}},
  \bibinfo{author}{\bibfnamefont{K.}~\bibnamefont{Tamasaku}},
  \bibinfo{author}{\bibfnamefont{N.}~\bibnamefont{Ichikawa}}, \bibnamefont{and}
  \bibinfo{author}{\bibfnamefont{S.}~\bibnamefont{Uchida}},
  \bibinfo{journal}{Phys. Rev. Lett.} \textbf{\bibinfo{volume}{77}},
  \bibinfo{pages}{5417} (\bibinfo{year}{1996}).

\bibitem[{\citenamefont{Vojta et~al.}(2000)\citenamefont{Vojta, Buragohain, and
  Sachdev}}]{vojt00}
\bibinfo{author}{\bibfnamefont{M.}~\bibnamefont{Vojta}},
  \bibinfo{author}{\bibfnamefont{C.}~\bibnamefont{Buragohain}},
  \bibnamefont{and} \bibinfo{author}{\bibfnamefont{S.}~\bibnamefont{Sachdev}},
  \bibinfo{journal}{Phys. Rev. B} \textbf{\bibinfo{volume}{61}},
  \bibinfo{pages}{15152} (\bibinfo{year}{2000}).

\bibitem[{\citenamefont{Xiao et~al.}(1990)\citenamefont{Xiao, Cieplak, and
  Chien}}]{xiao90}
\bibinfo{author}{\bibfnamefont{G.}~\bibnamefont{Xiao}},
  \bibinfo{author}{\bibfnamefont{M.~Z.} \bibnamefont{Cieplak}},
  \bibnamefont{and} \bibinfo{author}{\bibfnamefont{C.~L.} \bibnamefont{Chien}},
  \bibinfo{journal}{Phys. Rev. B} \textbf{\bibinfo{volume}{42}},
  \bibinfo{pages}{240} (\bibinfo{year}{1990}).

\end{thebibliography}

\end{document}